\documentclass{PoS}
\def\beq{\begin{equation}}   
\def\eeq{\end{equation}}
\def\bea{\begin{eqnarray}}  
\def\eea{\end{eqnarray}}

\def\ihixs{{\tt iHixs}}
\title{Precise inclusive Higgs predictions using iHixs}
\ShortTitle{Precise inclusive Higgs predictions using iHixs}

\author{\speaker{Stephan Buehler}\\
        ETH, Zurich\\
        E-mail: \email{buehler@itp.phys.ethz.ch}}
\abstract{
We present a new program ({\ihixs}) which computes the inclusive Higgs
boson cross section at hadron colliders. It  incorporates  QCD
corrections through NNLO, real and virtual electroweak corrections, 
mixed QCD-electroweak corrections, quark-mass effects through NLO in
QCD, and finite width effects for the Higgs boson and heavy
quarks. We focus in particular on the Higgs width effects and results using the 
recently added NNPDF21 parton distribution functions.\\
Cross section predictions are provided for the $8$ TeV LHC in the mass range from
$114$ to $131$ GeV.
}
\FullConference{10th International Symposium on Radiative Corrections (Applications of Quantum Field Theory to Phenomenology)\\
                 September 26-30, 2011\\
                 Mamallapuram, India}
\begin{document}
\section{Introduction}
For the last couple of decades, one of the main goals of the high energy physics community has been the discovery of the Higgs boson.
Huge efforts have been put forth on both the experimental and theoretical side. On the former, the LEP~\cite{Barate:2003sz} experiment at CERN, as well as the
TEVATRON~\cite{Aaltonen:2011gs} at Fermilab were able to constrain the mass range available for a Standard Model (SM) Higgs Boson.
With the LHC~\cite{Collaboration:2011qi,Chatrchyan:2011tz} performing impressively since the beginning last year, these exclusion limits are steadily improving.

To provide the best possible prediction for the Higgs boson production cross section and its uncertainties at hadron colliders, we have written a new computer program, called {\ihixs}
(inclusive Higgs cross sections). We include all known fixed-order contributions to the main Higgs boson production channel,
 gluon fusion, in a consistent way. To be able to accomodate beyond-the-Standard Model (BSM) studies including enhanced Yukawa couplings, we also 
added the usually subdominant bottom-quark fusion process to the program.
In addition, our goal was to have a proper treatment of the non-zero Higgs boson decay width, which becomes important for heavy Higgs masses, where the Higgs Resonance
grows broad.

{\ihixs} has been published in July 2011 and is presented in Ref.~\cite{Anastasiou:2011pi}. In this article, we will briefly list the components of {\ihixs} 
and then focus on studies concerning the Higgs width effects (section \ref{width}) and
the comparison of different parton density sets (section \ref{pdf}), including the sets of the NNPDF collaboration that were not present in the original publication.

In addition, in the light of the refined Higgs exclusion limits and signal-like excesses in the light Higgs mass region published by Atlas and CMS in December 2011~\cite{lhc},
we include a table of Higgs cross section predictions for the mass range from $114$ to $131$ GeV, assuming an $8$ TeV LHC. It can be found in section \ref{crosssection}.
\section{Components of iHixs}
{\ihixs} includes two Higgs boson production modes, gluon fusion and bottom-quark fusion. We will now briefly list each modes components. We kindly refer to section 2 of \cite{Anastasiou:2011pi}
for a more extensive overview, as well as thorough references that had to be omitted here due to lack of space.

The cross section for the gluon fusion process in {\ihixs} comprises
the LO and NLO QCD effects with exact quark-mass dependence,
the NNLO QCD corrections using heavy quark effective theory (HQET),
the two-loop electroweak corrections at LO in $\alpha$,
one-loop electroweak corrections to the real radiation processes $q \bar{q} \to g h$ and  $q g \to q h$,
as well as mixed QCD and electroweak contributions with light quarks.\\
The number and coupling strengths of the heavy quarks propagating in the loops is arbitrary, and the coupling of the electroweak gauge bosons to the Higgs boson can also be rescaled.
These features allow the prospective user to perform some BSM studies using {\ihixs}.

The cross section for bottom-quark fusion process in {\ihixs}  
comprises the LO, NLO and NNLO QCD effects. The Yukawa coupling to the Higgs boson can be rescaled by a factor (as in the gluon fusion case), allowing the study of BSM models that feature
enhanced bottom Yukawa couplings. An example of such a study can be found in section 8 of \cite{Anastasiou:2011pi}.
\section{Higgs width effects}
\label{width}
A light Higgs boson, as  predicted  in the Standard Model, has a
rather small $\delta\equiv \Gamma_H(m_H)/m_H$ and it  is often sufficient  to  take the
$\delta =0$ limit of the zero width approximation (ZWA).  
Existing experimental studies at  hadron
colliders~\cite{Aaltonen:2011gs,Chatrchyan:2011tz,Collaboration:2011qi}
have always  reported limits on the Higgs boson cross section
comparing with expectations in this approximation. 
 However, recent years have witnessed the alarming trend of using this approximation in situations where it may be 
insufficient.

The resonant part of the partonic cross section of initial partons $i$ and $j$ into a given final state $\{H_{final}\}$ (plus possible additionally radiated partons, collectively
denoted as $X$) can be written as
\begin{equation}
\label{eq:BW0}
\hat{\sigma}_{ij \to \{H_{\rm final}\}+X}(\hat{s}, \mu_f)    = 
\int_{Q_{a}^2}^{Q_{b}^2} 
dQ^2 \frac{Q \Gamma_H(Q)}{ \pi }
\frac{
 \hat{\sigma}_{ij \to H}(\hat{s}, Q^2, \mu_f) {\rm Br}_{H \to \{H_{\rm final} \}}(Q)
}
{ (Q^2 -m_H^2)^2} \, ,
\end{equation} 
where $Q_a,Q_b$ define the experimentally accessible range for the invariant mass of the final state and $\Gamma_H(Q)$ is the decay width of a Higgs boson at rest with mass $Q$.
 This expression diverges in the limit $Q\rightarrow m_H$, and a
 resummation of  resonant contributions at all orders is necessary in order to  render  the propagator finite in this limit.
We  remark that a resummation of partial perturbative  corrections  from
all perturbative orders into the propagator of an unstable particle
is a delicate theoretical issue~\cite{Beneke:2004km,Denner:2005fg}. 
Historically, it has been treated with various prescriptions in the
literature with varied success (see, for example, 
references in~\cite{Zanderighi:2004qu}).
To a  first approximation, the cross section becomes
\begin{equation}
\label{eq:BWG}
\hat{\sigma}_{ij \to \{H_{\rm final}\}+X}(\hat{s}, \mu_f)    = 
\int_{Q_{a}^2}^{Q_{b}^2} 
dQ^2 \frac{Q \Gamma_H(Q)}{ \pi }
\frac{
 \hat{\sigma}_{ij \to H}(\hat{s}, Q^2, \mu_f) {\rm Br}_{H \to \{H_{\rm final} \}}(Q)
}
{ (Q^2 -m_H^2)^2 + m_H^2 \Gamma_H^2(m_H)}.
\end{equation} 
We have implemented this integration over the Breit-Wigner (BW) distribution in {\ihixs} as the \verb default  option.
 Note that in eq. (\ref{eq:BWG}), we need the value of the Higgs width and the 
Branching ratio at every virtuality sampled by the integration. This should not be approximated with the respective values at $Q=m_H$
 since these values can be quickly changing as thresholds are being crossed.
For the Standard Model, these values are distributed in a grid file we have generated using the program \verb HDECAY  of Ref~\cite{Djouadi:1997yw}
. In order to use {\ihixs}  with an arbitrary BSM model, the user needs  to provide  a  data file  with the width
and branching ratios  of the Higgs  boson  as  a function  of the
virtuality of the Higgs boson.

In the Standard Model, it has been observed that significant
cancelations due  to interference of   resonant and  
non-resonant  diagrams  take place at high invariant  masses 
(Refs~\cite{Glover:1988rg,Glover:1988fe,Baur:1990mr,Valencia:1992ix}). 
{\ihixs}  takes into  account only diagrams with an s-channel Higgs  boson propagator.  
The  line-shape  away from the resonance is  therefore poorly described.  
To improve upon this, we  have implemented a prescription based on the
resummation of  
$VV \to VV$ scattering amplitudes  with  the  dominant contributions
from both resonant and  non-resonant 
Feynman diagrams at the  high energy regime. 
 Ref~\cite{Seymour:1995qg} 
performs  a Dyson re-summation of 
the tree-level Goldstone boson scattering amplitude  leading to an 
``improved  s-channel approximation'' (ISA). In this  framework, the Higgs
propagator is modified according to the prescription:
\begin{equation}
\label{eq:Seymprop}
\frac{i}{\hat{s}-m_H^2}  \to 
\frac{
i \frac{m_H^2}{\hat{s}}
}
{\hat{s}-m_H^2 +i \Gamma_H(m_H^2) \frac{\hat{s}}{m_H}},
\end{equation} 

which leads to the modified cross section
\begin{equation}
\label{eq:BWISA}
\hat{\sigma}_{ij \to \{H_{\rm final}\}+X}(\hat{s}, \mu_f)    = 
\int_{Q_{a}^2}^{Q_{b}^2} 
dQ^2 \frac{m_H^4 \Gamma_H(Q)}{Q^3 \pi }
\frac{
 \hat{\sigma}_{ij \to H}(\hat{s}, Q^2, \mu_f) {\rm Br}_{H \to \{H_{\rm final} \}}(Q)
}
{ (Q^2 -m_H^2)^2 + \frac{Q^4}{m_H^2} \Gamma_H^2(m_H)}.
\end{equation} 
It interpolates  smoothly between two limits which
are  well described either by resummation or by fixed-order
perturbation theory: the resonant region $Q \sim m_H$ and the high
energy limit $Q \gg m_H$. This scheme is also implemented in {\ihixs} and is used when the \verb Seymour  option is chosen.

\begin{figure}[htp]
 \centering
 \includegraphics[width=.7\textwidth]{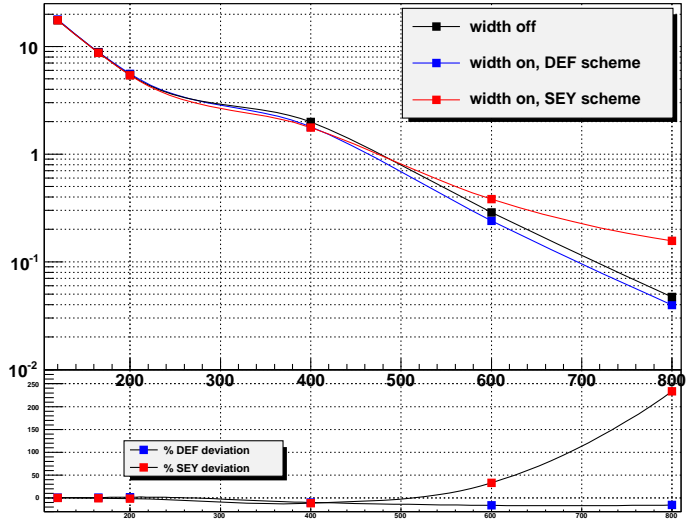}
 \caption{Comparison of the total cross section in the zero width approximation, $\sigma^{ZWA}$, with a finite width in the default scheme,
 $\sigma^{DEF}$ and in the Seymour scheme, $\sigma^{SEY}$. In the lower panel we show the relative error one makes when adopting the ZWA, 
defined as $\frac{\sigma-\sigma^{ZWA}}{\sigma^{ZWA}}\cdot100\%$.}
\label{fig3}
\end{figure}

The numerical impact of the scheme choice on the inclusive Higgs production cross section can be seen in figure \ref{fig3}. It shows the cross section 
as a function of the Higgs mass for the three possible scheme choices (ZWA, \verb default, \verb Seymour ), as well as their ratios (in the lower panel).
 For small Higgs masses, as expected, the differences are minute. The relative difference of the ZWA and the BW-integration reaches the percent level at
$m_H\sim150$ GeV, while the ISA stays closer to the ZWA. Above $m_H\sim500$ GeV, the ISA starts to deviate enourmously from the other
schemes and predicts a much larger cross section due to the signal-background interference it tries to simulate.

\begin{figure}[hbtp]
\centering
\begin{minipage}[b]{0.49\textwidth}
\includegraphics[width=\textwidth]{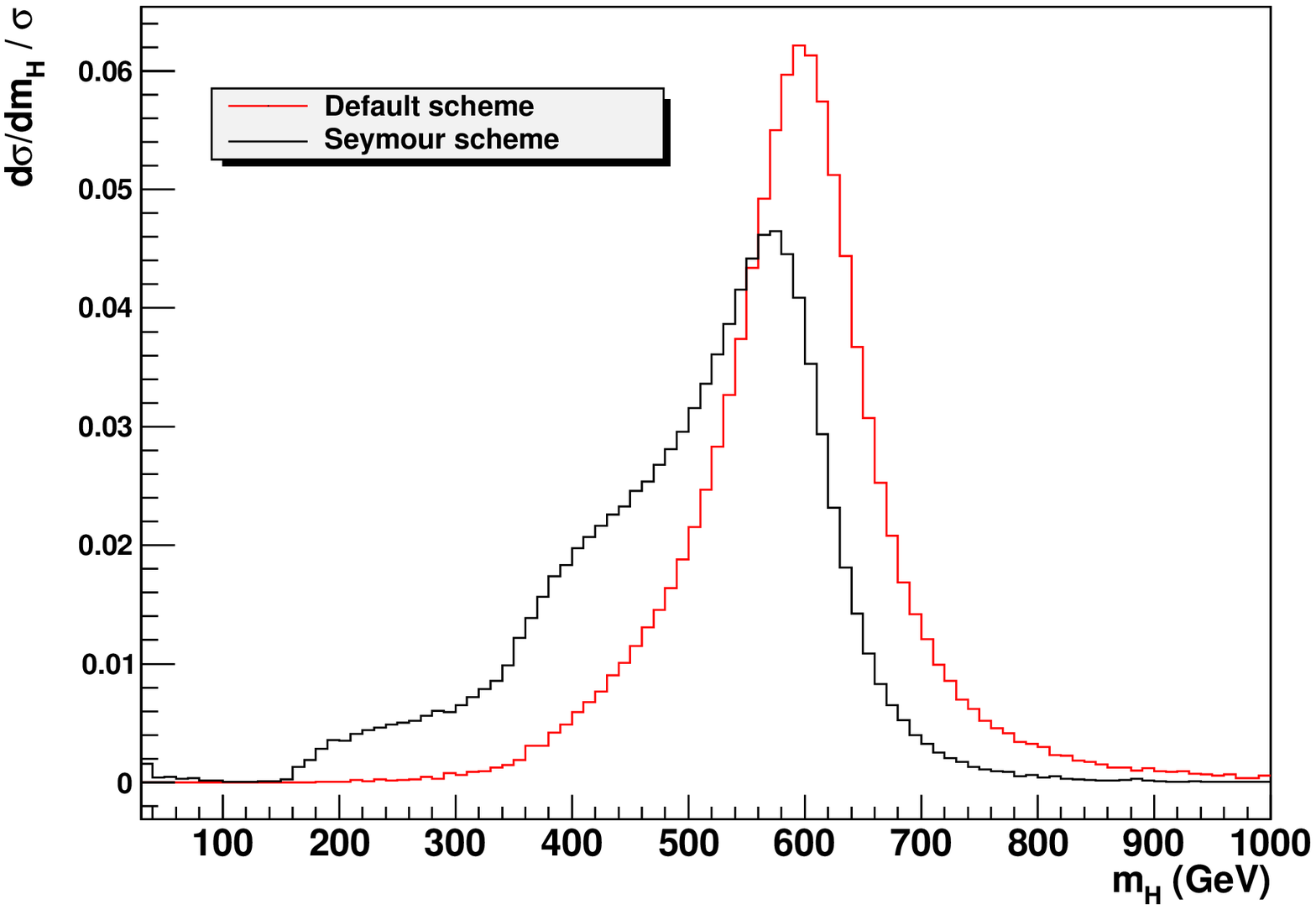}
\end{minipage}
\begin{minipage}[b]{0.49\textwidth}
\includegraphics[width=\textwidth]{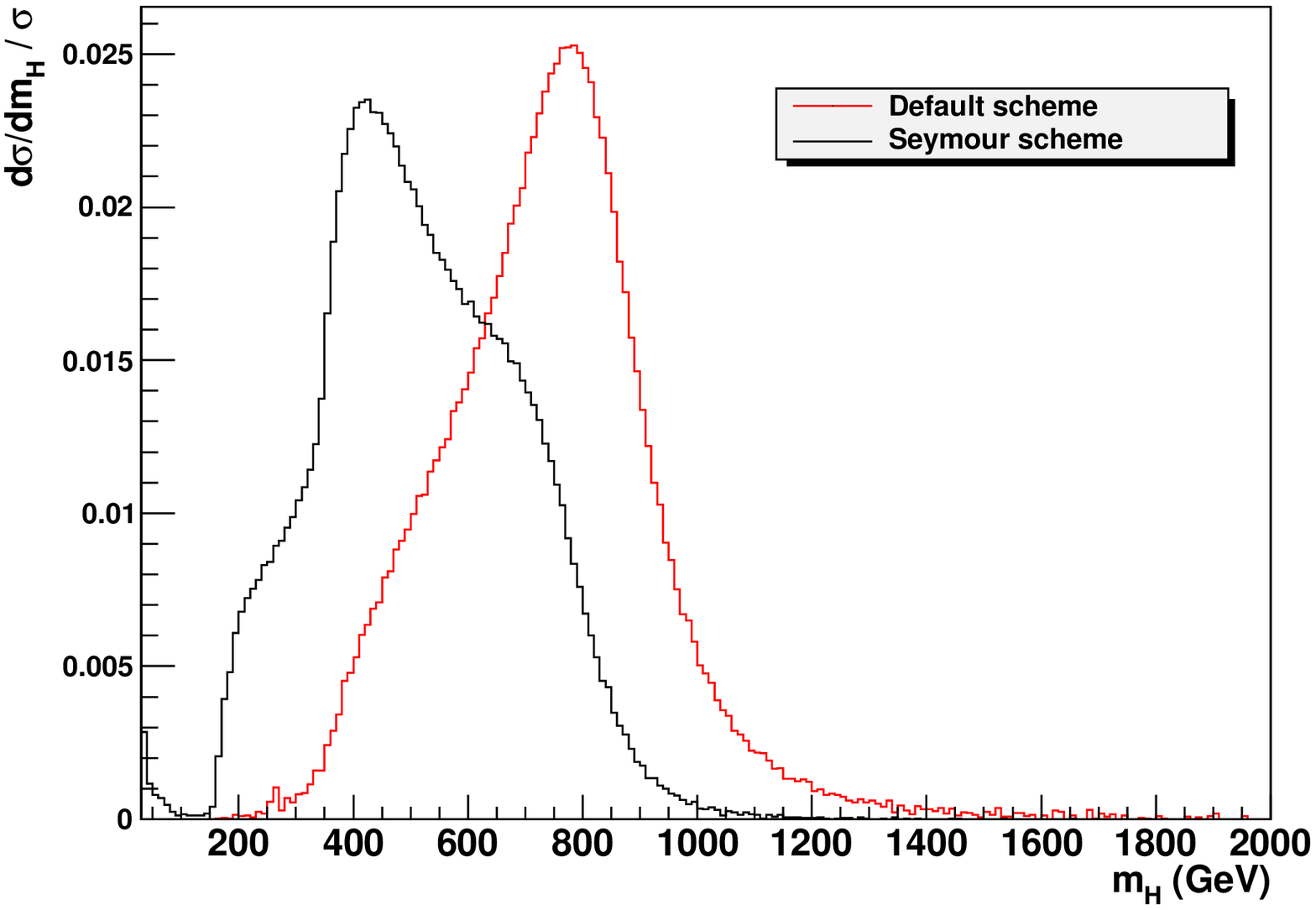}
\end{minipage}
\caption{The invariant mass distribution of the Higgs boson with $m_H=600$, $800$ GeV, in the default and the ISA scheme.}
\label{invmassdistro}
\end{figure}

Within this context, it is interesting to notice that the invariant mass distribution of the Higgs boson, shown in figure \ref{invmassdistro},
 gets significantly distorted in the high mass region, where the Higgs width is large. The distortion is spectacularly stronger in the case of
 the ISA, as a consequence of the fact that the scheme tries to simulate the effects of signal-background interference off the resonant peak. 
These effects become increasingly important for high Higgs masses.

In a recent publication~\cite{Campbell:2011cu}, the cross section including signal-background interference at LO in QCD 
for the process $gg\rightarrow H \rightarrow WW$ has been compared with the improved s-channel approximation. The authors found that the
 ISA indeed captures some features of the interference. At differential level, though, they find the shapes of invariant mass distributions to differ significantly.

We  believe that the Higgs boson line-shape will enjoy many future
theoretical studies with improved resummation methods for resonant
diagrams and matching to fixed-order  perturbation theory away from
the resonance region.

\section{Parton density function comparison}
\label{pdf}
\begin{figure}[htbp]
\centering
\begin{minipage}[t]{0.49\textwidth}
 \includegraphics[angle=-90,width=\textwidth]{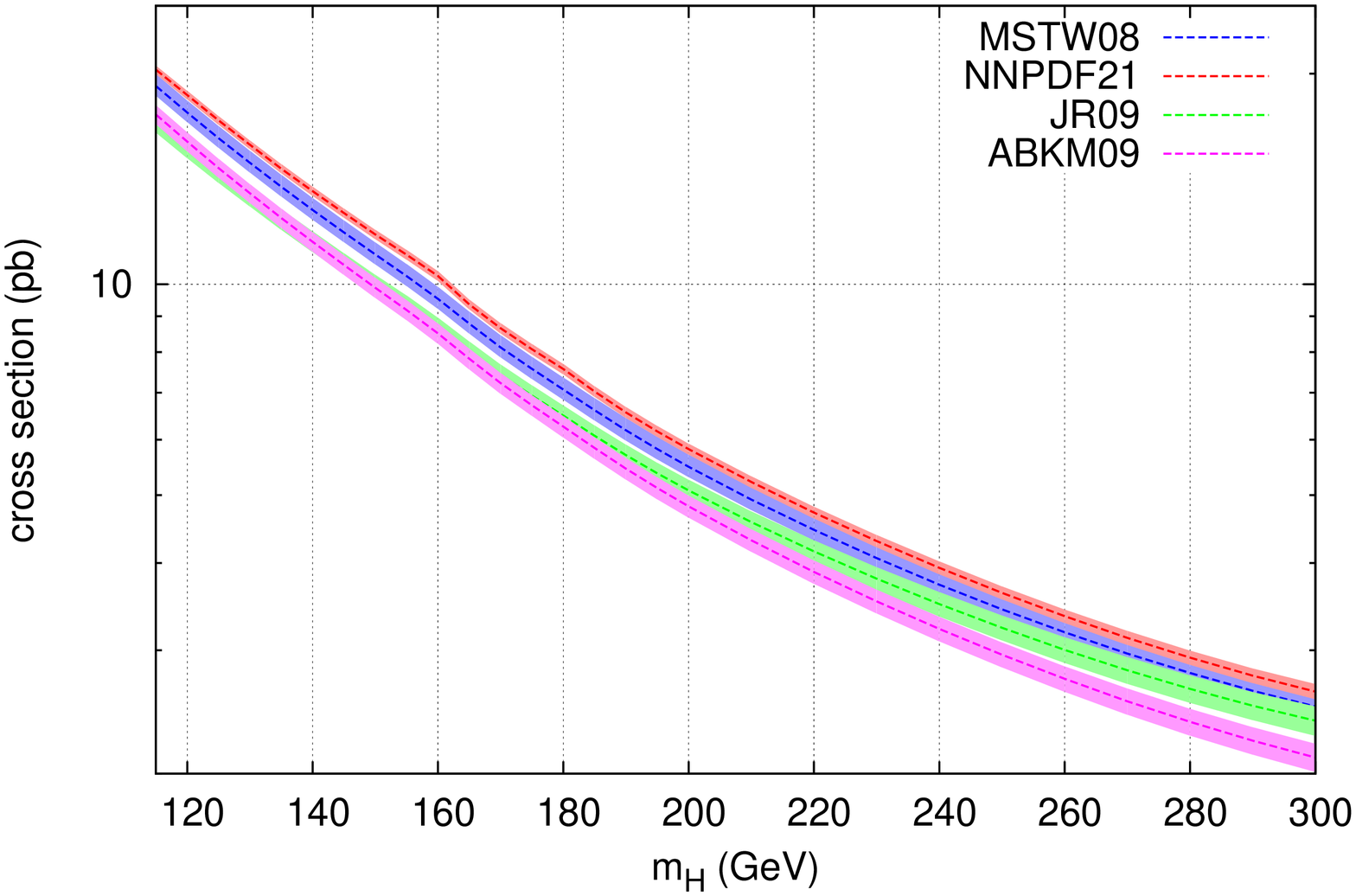}
\end{minipage}
\begin{minipage}[t]{0.49\textwidth}
 \includegraphics[angle=-90,width=\textwidth]{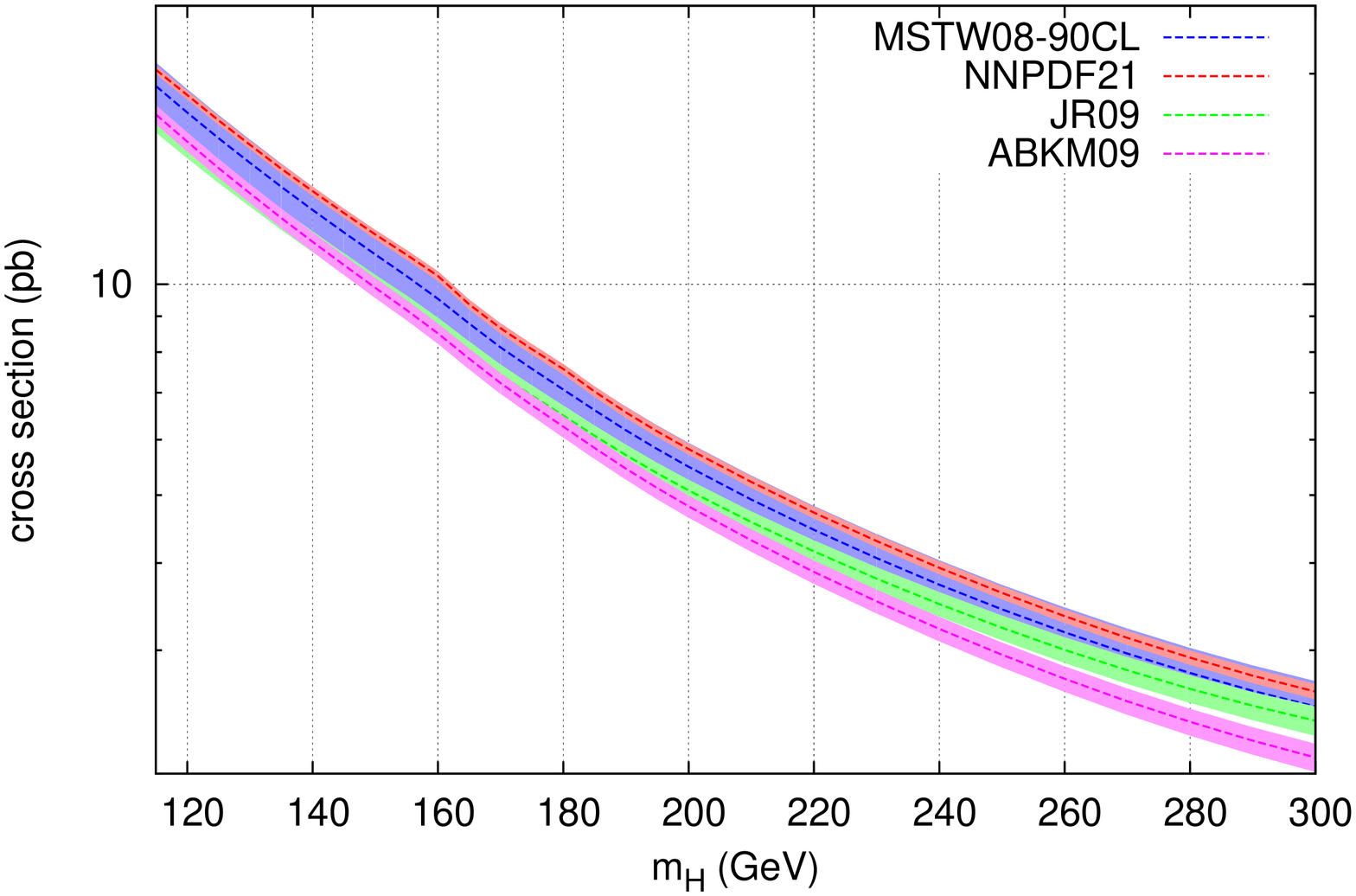}
\end{minipage}
\caption{The gluon fusion cross section as a function of the Higgs boson mass, for all PDF providers supported in {\ihixs}. The LHS shows the $68\%CL$ uncertainty
bands, while on the RHS, the MSTW band corresponds to the $90\%CL$ uncertainty.}
\label{fig2}
\end{figure}

{\ihixs}  computes the inclusive  Higgs boson cross section through  NNLO in perturbative QCD. 
A  large  source  of uncertainty for Higgs  boson cross sections at hadron colliders  is  the precision in the  determination of the parton densitiy functions (PDFs).
It is  therefore important to compare the effect of  diverse existing determinations of parton densities  on the  Higgs cross section, as 
well as future  sets  which will incorporate refined measurements  and  theory. 
{\ihixs}  allows these studies  effortlessly. It is interfaced  through the  LHAPDF library~\cite{Whalley:2005nh} 
with all available parton distribution functions with a  consistent  evolution at NNLO.~\cite{Alekhin:2009ni,Martin:2009iq,JimenezDelgado:2009tv,Ball:2011uy}. 
The last set, the first NNLO set provided by the NNPDF collaboration, has only recently been published and was not present in our original
publication~\cite{Anastasiou:2011pi}. We present updated plots comparing the different PDF sets in figure \ref{fig2}.

Figure \ref{fig2} shows the gluon fusion cross section in the mass range between $115$ and $300$ GeV, for all the PDF providers supported by ihixs (ABKM09, JR09, MSTW08, NNPDF21).
All the curves are engulfed in their $68\%CL$ uncertainty bands that were calculated according to each providers prescription. The MSTW collaboration
also provides sets to estimate the uncertainty to $90\%CL$, which is shown on the right hand side of figure \ref{fig2}.\\
Clearly, the NNPDF band is more narrow than the others. This is due to the fact that until now,
there is only one NNLO set available by the NNPDF collaboration in the LHAPDF interface, which comes with a fixed value for $\alpha_s(m_Z)$. The other bands, on the other hand, incorporate the combined
$\alpha_s$+PDF error. Since the gluon fusion cross section is $\mathcal{O}(\alpha_s^2)$ already at LO, this enhances the uncertainty significantly.\\
We observe that, while the NNPDF curve is above all the other curves throughout. Its error band overlaps with the MSTW one except for a region around $160$ GeV.
On the right hand side, the bands overlap over the whole mass range.
The cross section predictions by ABKM and JR are below the MSTW ones throughout,
with the JR curve behaving slightly different than the other three. Furthermore, there is an obvious gap between the respective error bands, even when using the
$90\%CL$ sets from MSTW. This suggests that when adopting one provider and its precscription, one possibly underestimates the PDF uncertainty.\\
Parts of this discrepancy can be explained by the different values of $\alpha_s(m_Z)$ that
each provider chooses, since the order of curves in the low-mass region is the same as the order of the $\alpha_s(m_Z)$ values adopted.
\section{Cross section prediction in the mass range $114$ to $131$ GeV}
\label{crosssection}
On December 12, 2011, the Atlas and CMS collaborations published their updated results concerning the SM Higgs search, using the full 2011 dataset~\cite{lhc}.
The experiments are now able to exclude the SM Higgs boson at $95\%CL$ from $131$ GeV ($127$ for CMS) up to very high Higgs masses around $600$ GeV. On top of that,
there is an excess of signal-like events for invariant masses around $125$ GeV that appears in multiple channels and both experiments. Due to this recent development, we present in
table \ref{table1} our best prediction for the Higgs production cross section in gluon fusion in the mass range that is not excluded yet. They include scale- and PDF-uncertainties and 
assume a hadronic center-of-mass energy of $8$ TeV. The central scales used are $\mu_F=\mu_R=m_H/2$ and scale variations are determined by varying these simultaneously by a factor of two. The PDF set 
that was used is the MSTW08 set.
\begin{table}
\centering
  \begin{tabular}{| c ||  c | c | c |c|c|c|}
\hline
$m_H$&$\sigma(pb)$ & $\% \delta_{PDF}^+$ & $\% \delta_{PDF}^-$ & $\% \delta_{\mu}^-$ & $\% \delta_{\mu}^+$\\ 
\hline\hline
114.0&	24.69&	4.00&	-3.04&	8.83&	-9.32\\ \hline
115.0&	24.27&	3.99&	-3.04&	9.09&	-9.31\\ \hline
116.0&	23.94&	3.98&	-3.07&	8.75&	-9.60\\ \hline
117.0&	23.55&	4.00&	-3.05&	8.66&	-9.33\\ \hline
118.0&	23.17&	3.99&	-3.05&	8.61&	-9.39\\ \hline
119.0&	22.79&	4.00&	-3.05&	8.57&	-9.35\\ \hline
120.0&	22.42&	3.99&	-3.05&	8.55&	-9.31\\ \hline
121.0&	22.05&	3.99&	-3.05&	8.54&	-9.30\\ \hline
122.0&	21.70&	3.99&	-3.04&	8.50&	-9.27\\ \hline
123.0&	21.36&	3.98&	-3.04&	8.45&	-9.28\\ \hline
124.0&	21.02&	3.99&	-3.02&	8.42&	-9.25\\ \hline
125.0&	20.69&	3.98&	-3.06&	8.37&	-9.26\\ \hline
126.0&	20.37&	3.97&	-3.07&	8.36&	-9.24\\ \hline
127.0&	20.05&	3.98&	-3.05&	8.35&	-9.21\\ \hline
128.0&	19.74&	3.98&	-3.05&	8.32&	-9.20\\ \hline
129.0&	19.44&	3.99&	-3.04&	8.29&	-9.26\\ \hline
130.0&	19.14&	3.98&	-3.05&	8.26&	-9.19\\ \hline
131.0&	18.85&	3.98&	-3.04&	8.24&	-9.16\\ \hline

 \end{tabular}
 \caption{Total cross section for LHC at $\sqrt{s}=8$ TeV with MSTW PDF errors (corresponding to $68\%$CL).}
\label{table1}
\end{table}
\section{Conclusions}
We have presented our computer program, {\ihixs}, that calculates the inclusive Higgs boson production cross section both in gluon fusion and bottom-quark fusion.
{\ihixs}  provides the most precise predictions for
the Higgs boson rate at hadron colliders in  fixed order perturbation
theory, including QCD corrections through NNLO and electroweak
corrections for virtual and real radiative partonic processes.  

Using {\ihixs}, we have performed a study on the influence of a nonzero Higgs boson decay width on cross section predictions. We have found that, while effects are
small in the low-mass regime, there are significant changes for a heavy Higgs boson. Furthermore, the results are strongly scheme dependent, mirroring our lack of understanding
such broad resonances.

We have also presented an updated comparison among the different NNLO parton distribution functions, now including the recently published NNPDF21 NNLO set.
We have found large differences in their respective cross section predictions, with the $68\%CL$ uncertainty bands not overlapping for every pair of providers.
This is suspected to be largely due to different choices of the strong coupling constant.

Lastly, we have presented our best prediction for gluon fusion Higgs production at the $8$ TeV LHC in the mass range that is not excluded yet by the CMS and Atlas
experiments.

For more details on {\ihixs} and extensive references on the subject of Higgs production, as well as instructions on how to download and run the code, we kindly
refer to our original publication \cite{Anastasiou:2011pi}.
\acknowledgments
We would like to thank the organisers of the RADCOR 2011 conference for the invitation to India and their hospitality.
Research supported by the Swiss National Foundation under contract SNF 200020-126632.

\end{document}